\input harvmac
\input epsf.sty


\noblackbox

\font\cmss=cmss10
\font\cmsss=cmss10 at 7pt

\def\inbar{\vrule height1.5ex width.4pt depth0pt}

\def\IN{\relax{\rm I\kern-.18em N}}
\def\IB{\relax\hbox{$\inbar\kern-.3em{\rm B}$}}
\def\IC{\relax\hbox{$\inbar\kern-.3em{\rm C}$}}
\def\IQ{\relax\hbox{$\inbar\kern-.3em{\rm Q}$}}
\def\ID{\relax\hbox{$\inbar\kern-.3em{\rm D}$}}
\def\IE{\relax\hbox{$\inbar\kern-.3em{\rm E}$}}
\def\IF{\relax\hbox{$\inbar\kern-.3em{\rm F}$}}
\def\IG{\relax\hbox{$\inbar\kern-.3em{\rm G}$}}
\def\IGa{\relax\hbox{${\rm I}\kern-.18em\Gamma$}}
\def\IH{\relax{\rm I\kern-.18em H}}
\def\IK{\relax{\rm I\kern-.18em K}}
\def\IL{\relax{\rm I\kern-.18em L}}
\def\IP{\relax{\rm I\kern-.18em P}}
\def\IR{\relax{\rm I\kern-.18em R}}
\def\Z{\relax\ifmmode\mathchoice{\hbox{\cmss Z\kern-.4em Z}}{\hbox{\cmss Z\kern-.4em Z}} {\lower.9pt\hbox{\cmsss Z\kern-.4em Z}}{\lower1.2pt\hbox{\cmsss Z\kern-.4em Z}}\else{\cmss Z\kern-.4em Z}\fi}

\def\II{\relax{\rm I\kern-.18em I}}
\def\one{\relax{\rm 1\kern-.25em I}}

\def\S{Sec.~}

\def\CLL{\relax{\CL\kern-.74em \CL}}
\def\p{{\partial}}

\def\CM{{\cal M}}

\lref\john{
  D.~T.~Son,
  arXiv:0804.3972 [hep-th];\
  K.~Balasubramanian and J.~McGreevy,
  arXiv:0804.4053 [hep-th];\
  W. Goldberger, arXiv:0806.2867 [hep-th];\
  J. Barbon and C. Fuertes, arXiv:0806.3244 [hep-th];\
  W. Wen, arXiv:0807.0633 [hep-th];
  C. Herzog, M. Rangamani and S. Ross, arXiv:0807.1099 [hep-th];\
J. Maldacena, D. Martelli and Y. Tachikawa, arXiv:0807.1100 [hep-th];\
A. Adams, K. Balasubramanian and J. McGreevy, arXiv:0807.1111 [hep-th];
D. Minic and M. Pleimling, arXiv:0807.3665 [cond-mat];
J. Chen and W. Wen, arXiv:0808.0399 [hep-th].
}

\lref\Fradkin{E. Ardonne, P. Fendley and E. Fradkin, Annals. Phys. {\bf 310} (2004) 493 [arXiv:cond-mat/0311466].}

\lref\Greg{J. Maldacena, G. Moore and N. Seiberg, JHEP {\bf 0110} (2001) 005 [arXiv:hep-th/0108152].}

\lref\RK{D. Rokhsar and S. Kivelson, Phys. Rev. Lett. {\bf 61} (1988) 2376.}

\lref\Henley{C.L. Henley, J. Stat. Phys. {\bf 89} (1997) 483.}

\lref\Grinstein{G. Grinstein, Phys. Rev. {\bf B23} (1981) 4615.}

\lref\HLS{R. Hornreich, M. Luban and S. Shtrikman, Phys. Rev. Lett. {\bf 35} (1975) 1678.}

\lref\CM{C. Herzog, P. Kovtun, S. Sachdev and D.T. Son, Phys. Rev.
{\bf D75} (2007) 085020 [arXiv:hep-th/0701036];
A. Karch and A. O'Bannon, JHEP {\bf 0709} (2007) 024 [arXiv:0705.3870 [hep-th]];
S. Hartnoll, P. Kovtun, M. Muller and S. Sachdev, Phys. Rev. {\bf B76} (2007)
144502 [arXiv:0706.3215 [cond-mat]];
S. Hartnoll and C. Herzog, Phys. Rev. {\bf D76} (2007) 106012
[arXiv:0706.3228 [hep-th]];
S. Hartnoll and C. Herzog, Phys. Rev. {\bf D77} (2008) 106009
[arXiv:0801.1693[hep-th]];
S. Gubser, arXiv:0801.2977 [hep-th];
S. Gubser and S. Pufu, arXiv:0805.2960 [hep-th];
S. Hartnoll, C. Herzog and G. Horowitz, arXiv:0803.3295 [hep-th];
M. Roberts and S. Hartnoll, arXiv:0805.3898 [hep-th];
E. Keski-Vakkuri and P. Kraus, arXiv:0805.4643[hep-th];
A. Karch, D.T. Son and A.O. Starinets, arXiv:0806.3796 [hep-th].
S. Gubser and F. Rocha, arXiv:0807.1737 [hep-th].
}

\lref\FHMOS{E. Fradkin, D. Huse, R. Moessner, V. Oganesyan and S. Sondhi, Phys. Rev. {\bf B69} (2004) 224415.}

\lref\VBS{A. Vishwanath, L. Balents and T. Senthil, Phys. Rev. {\bf B69} (2004) 224416.}

\lref\GVS{P. Ghaemi, A. Vishwanath and T. Senthil, Phys. Rev. {\bf B72} (2005) 024420.}

\lref\Chetan{
  M.~Freedman, C.~Nayak, and K.~Shtengel,
  Phys. Rev. Lett. {\bf 94} (2005) 147205 [arXiv: cond-mat/0408257].
}

\lref\Yang{K. Yang, Phys. Rev. Lett. {\bf 93} (2004) 066401.}

\lref\SS{S. Sachdev and T. Senthil, Ann. Phys. {\bf 251} (1996) 76.}

\lref\Sachdev{S. Sachdev, {\it Quantum~Phase~Transitions}, Cambridge University Press, 1999.}

\lref\GubserBC{
  S.~S.~Gubser, I.~R.~Klebanov and A.~M.~Polyakov,
  Phys.\ Lett.\  B {\bf 428}, 105 (1998)
  [arXiv:hep-th/9802109].
}

\lref\Witten{E. Witten, Adv. Th. Math. Phys. {\bf 2} (1998) 253.}

\lref\Juan{J. Maldacena, Adv. Th. Math. Phys. {\bf 2} (1998) 231.}

\lref\Gary{G.T. Horowitz and S.F. Ross, Phys. Rev. {\bf D56} (1997) 2180 [arXiv:hep-th/9704058].}

\lref\BianchiKW{
  M.~Bianchi, D.~Z.~Freedman and K.~Skenderis,
  Nucl.\ Phys.\  B {\bf 631}, 159 (2002)
  [arXiv:hep-th/0112119].
}

\lref\KlebanovTB{
I.~R.~Klebanov and E.~Witten,  
Nucl.\ Phys.\  B {\bf 556}, 89 (1999)  [arXiv:hep-th/9905104].  
}

\lref\Varma{C.M. Varma, Phys. Rev. {\bf B55} (1997) 14554.}

\lref\Si{Q. Si et al, Nature {\bf 413} (2001) 804; Phys. Rev. {\bf B68} (2003) 115103.}

\lref\BreitenlohnerBM{P. Breitenlohner and D.Z. Freedman,
Phys. Lett. {\bf B115} (1982) 197.}

\Title{\vbox{\baselineskip12pt\hbox{SLAC-PUB-13354, SU-ITP-08/16}
\hbox{} }}
{\vbox{ {\centerline{Gravity Duals of Lifshitz-like Fixed Points} }}}

\centerline{Shamit Kachru,$^{a}$ Xiao Liu,$^{b}$ and Michael Mulligan$^{a}$}
\bigskip
{\it \centerline{$^{a}$Department of Physics and SLAC, Stanford University, Stanford, CA 94305, USA}}
{\it \centerline{$^{b}$Perimeter Institute for Theoretical Physics, Waterloo, Ontario, N2L 2Y5, Canada}}

\bigskip
\noindent

We find candidate macroscopic gravity duals for scale-invariant but non-Lorentz invariant fixed points,
which do not have particle number as a conserved quantity. 
We compute two-point correlation functions which exhibit novel behavior relative to their AdS counterparts,
and find holographic renormalization group flows to conformal field theories.
Our 
theories are characterized by a dynamical critical exponent $z$, which governs the anisotropy between spatial and temporal scaling $t \to \lambda^z t$, $x \to \lambda x$; we focus on the case with $z=2$.
Such theories describe multicritical points in certain magnetic materials and liquid
crystals, and have been shown to arise at quantum critical points in toy models
of the cuprate superconductors.  This work can be considered a small step towards making useful
dual descriptions of such critical points.

\Date{August, 2008}

\newsec{Introduction and Motivation}

In many condensed matter systems, one finds phase transitions governed by fixed points which exhibit ``dynamical scaling"
\eqn\dyn{t \to \lambda^{z}t, ~x\to \lambda x,~z \neq 1}
instead of the more familiar scale invariance which arises in the conformal group
\eqn\con{t \to \lambda t,~x \to \lambda x~.}
A toy model which exhibits this scale invariance (and which is analogous in many ways to the free scalar field example
of a standard conformal field theory) is the Lifshitz field theory:
\eqn\lif{{\cal L} = \int~d^2x~dt~\left( (\partial_t \phi)^2 - \kappa (\nabla^2 \phi)^2 \right)~.}
This theory has a line of fixed points parametrized by $\kappa$ \Grinstein\ and arises at finite temperature
multicritical points
in the phase diagrams of known materials
\refs{\HLS,\Grinstein}.
It enjoys the anisotropic scale invariance \dyn\ with $z=2$.

This fixed point and its interacting cousins have become a subject of renewed interest in the context of strongly correlated
electron systems.  For instance, in the Rokhsar-Kivelson dimer model \RK, there is a zero-temperature quantum critical point
which lies in the universality class of \lif\ \Henley\ (for a nice general exposition of the importance of quantum critical points,
see \Sachdev).  Similar critical points also arise in more general lattice models of
strongly correlated electrons \refs{\FHMOS,\VBS,\Fradkin}.
The correlation functions in these models have interesting properties like finite-temperature ultra-locality in space at fixed time \GVS,
which may be important in explaining certain experimental results \refs{\Varma,\Si}.  Such theories
are also of interest in 1+1 dimensional systems \refs{\SS,\Yang}.

Furthermore, such fixed points seem to have a non-trivial generalization to non-Abelian gauge theories.
The Lagrangian \lif\ can be dualized to that for an Abelian gauge field in a standard way, since scalars are dual to vectors in 2+1
dimensions.  This yields a Lagrangian with the unusual property that the usual ${\bf E}^2$ term has vanishing coefficient; the leading
terms in the Lagrangian depend on derivatives 
of the electric field.  Freedman, Nayak and Shtengel analyzed a similar
theory with $SU(2)$ gauge group in \Chetan.  They provided evidence that the $SU(2)$ theory as well, has an interacting line of
fixed points with $z=2$.  However, these fixed points are strongly coupled; their existence and their detailed properties (correlation
functions, etc.) lie outside of the regime where the analysis in \Chetan\ was performed most reliably.

AdS/CFT duality \refs{\Juan,\GubserBC,\Witten} provides a well known technique to obtain weakly coupled and calculable dual descriptions
of strongly-coupled conformal theories, in terms of gravity or string theory on a weakly curved spacetime.  The symmetries of the
gravitational background geometrically realize the symmetries of the dual field theory; so for instance the conformal group
$SO(D,2)$ of a D-dimensional CFT, arises as the group of isometries of $AdS_{D+1}$.  It is natural to ask: can we find a more general
class of spacetimes which could be dual to theories with non-trivial dynamical critical exponents as in \dyn?
The general such theory has far fewer symmetries than a conformal theory; it enjoys scale invariance, invariance under spatial
and temporal translations, spatial rotation invariance, and P and T symmetry.\foot{Non-relativistic theories without particle production and with
dynamical scaling enjoy a larger symmetry group. When $z=2$, the enhanced symmetry group is called the Schr\"odinger group.
Gravity duals for such theories have
been studied in the recent works \john, initiated by the papers of Son and of Balasubramanian and McGreevy.
One potential application seems to be to systems of cold atoms at the unitarity limit.
The theories we study, while also lacking Lorentz invariance, have particle production; their structure and their potential
applications are quite different from those in \john.
Other recent papers applying AdS/CFT duality to study
different problems of
condensed matter physics in a similar spirit have appeared in e.g. \CM.
}

In this paper, we attack this question.  While we will focus most concretely on the case $z=2$ (where many of the most interesting
examples arise in the condensed matter literature), the techniques we use are clearly more general.  In \S 2, we show that 4D gravity with
a negative cosmological term, in the presence of a modest set of $p$-form gauge fields, can support metrics
which geometrize the symmetries of these Lifshitz-like fixed points.
In \S3, we compute two-point correlation functions for the simplest scaling operators, those dual to free bulk scalar fields.
Unlike the situation in conformal field theories, these correlators contain more information than can be inferred
from the scaling dimension of the operator alone.
We study the renormalization group flows between
our fixed points and conventional conformal field theories (analogous to the flow that would arise by perturbing the Lagrangian
\lif\ by the operator $- (\nabla\phi)^2$) in \S4.
Finally, we conclude with a discussion of several interesting questions that we hope to address in the future.

\newsec{Gravitational Solutions}

In this section, we find gravity solutions which have the right properties to be dual to (interacting generalizations of) Lifshitz
fixed points.  We work with general $z$ here, but specialize to $z=2$ in the next two sections when we compute correlation
functions and study renormalization group flows.

\subsec{The Metric}

We would like a metric invariant under the modified scale transformation \dyn.  In addition, we wish to study field theories
invariant under time and space translations, spatial rotations, spatial parity, and time-reversal.
We assume, as in AdS/CFT, that the ``scale" in the dual field theory is geometrized by the presence of an additional radial dimension
on the gravity side of the duality, and that rescaling of this radial coordinate geometrizes the scale transformations \dyn.

These assumptions lead us to the family of geometries (one for each value of $z$):
\eqn\metric{ds^2 = L^2 \left( - r^{2z} dt^2 + r^2 d{\bf x}^2 + {dr^2 \over r^2} \right),}
where $0<r<\infty$, $d{\bf x}^{2} = dx_{1}^{2} + ... + dx_{d}^{2}$, and $L$ sets the scale for the radius of curvature of the geometry. We set
the Planck length $l_{\rm pl}=1$, and hence every quantity above is dimensionless. The scale transformation acts as
\eqn\scaling{t \to \lambda^{z}t,~x \to \lambda x,~r \to {r\over \lambda}~.}
$z=1$ gives the usual metric on $AdS_{d+2}$, which enjoys the larger symmetry $SO(d+1,2)$.

This metric is nonsingular. All local invariants
constructed from the Riemann tensor remain finite everywhere, and in fact are constant. The latter fact follows from the symmetries of the geometry.
The Ricci scalar takes the value $- 2(z^2 + 2z + 3)/L^{2}$. The metric is not geodesically complete and has peculiar behavior near $r=0$; for $z \neq 1$, although the curvature invariants are small for sufficiently
large $L$, an infalling object experiences very large tidal forces.  This behavior is familiar from the
solutions corresponding to various string theoretic brane systems \Gary.   The lesson from these systems is that a metric like \metric\ is physically sensible if there exists a regular black hole solution that approaches it in an extremal limit;
we leave the construction of such black holes to future work.\foot{We are grateful to Sean Hartnoll and Gary Horowitz for bringing this subtlety to our attention.}

Henceforth, we will focus on the case $d=2$ (appropriate for gravitational duals to 2+1 dimensional field theories); so we will be studying
gravity in four dimensions.

\subsec{Full Solution}

We would like to obtain the metrics \metric\ as solutions of the field equations of General Relativity coupled to some matter content (which can presumably
arise in the low-energy limit of string or M-theory).
Because our theory, like the toy model \lif, may be expected to flow to normal CFTs under relevant perturbations, a good starting point will be
gravity with a negative cosmological term; this will be able to capture the end of any such RG flows via AdS/CFT duality.

Einstein gravity with a negative cosmological constant alone does not support the metrics \metric.  However, string theory also generically gives rise
to p-form gauge fields.  A modest choice of such additional content, which can support the metrics \metric, involves the addition of gauge fields
with $p=1,2$.

The Lagrangian is then given by:
\eqn\action{\ S =  \int d^4 x \sqrt{-g} \left(R - 2 \Lambda\right) - {1\over 2} \int  \left( {1\over e^2} F_{(2)} \wedge \ast F_{(2)} + F_{(3)} \wedge \ast F_{(3)} \right)  - c \int B_{(2)} \wedge F_{(2)}. }
where $F_{(2),(3)}$ are the field strengths for the gauge fields, $F_{(2)} = dA_{(1)}, F_{(3)} = dB_{(2)}$.
In addition to the standard Einstein-Hilbert action, a 4D cosmological constant $\Lambda$, and the kinetic terms for the gauge fields, we introduced
a topological coupling between the two and three form fluxes $\int B_{(2)} \wedge F_{(2)} = \int A_{(1)} \wedge F_{(3)}$ (up to boundary terms).
This coupling is necessary to find appropriate solutions of the equations of motion.
The topological coupling $c$ needs to be quantized in many (but not all conceivable) cases, as discussed in detail in for instance appendix A of \Greg.
However, note that after redefining $F_{(2)} \to {1\over e} F_{(2)}$, the action can be written with no $e$ in front of the gauge kinetic term, but
with $c \to ce$; in this way, we can consider $c$ to be arbitrarily small (at weak gauge coupling), and set $e=1$.  We proceed with this convention
henceforth.

To source the metric \metric, we need to turn on background two and three-form fluxes that preserve the symmetries:
\eqn\fluxes{F_{(2)} = A ~\theta_r \wedge \theta_t,~~F_{(3)} = B~\theta_r \wedge \theta_x \wedge \theta_y~.}
Here we work with a non-coordinate basis for the one-forms:
\eqn\basis{\theta_t = L r^z dt,~\theta_{x^i} = L r dx^i,~\theta_{r} = L {dr\over r}~.}
In this basis, the metric simplifies to ${\rm diag}(-1,1,1,1)$.  We fix our convention to be $\epsilon_{\theta_r \theta_t \theta_x \theta_y}=1$.  Both
fluxes are then closed.  Their field equations
\eqn\gaugeqn{d * F_{(2)} = -c F_{(3)},~d* F_{(3)} = c F_{(2)}}
then fix the value of the dynamical exponent $z$ in terms of the topological coupling $c$ and the radius of curvature $L$ of the dual geometry:
\eqn\zis{2z = (cL)^2~.}
The ratio between the two fluxes is also fixed in terms of the dynamical exponent:
\eqn\ratio{{A\over B} = \sqrt{{z \over 2}}~.}
Note that for small $c$, we can obtain a weakly curved geometry with any desired value of $z$.

The Einstein equation for the action \action
\eqn\einsteingen{G_{\mu \nu} + \Lambda g_{\mu \nu} = \sum_{p=2,3} {1 \over 2 p!} \left(p F_{\mu \rho_2 \cdots \rho_p} F_{\nu}^{\rho_2 \cdots \rho_p} - {1 \over 2} g_{\mu \nu} F_{\rho_1 \cdots \rho_p} F^{\rho_1 \cdots \rho_p} \right)}
then fully determines the required values of the background fluxes and the cosmological constant. As a result of the
symmetry of \metric\ and \fluxes, and the Bianchi identity, there are only two independent equations from the Einstein equations, which can be taken to be the tt and xx components:
\eqn\einstein{\eqalign { {z^{2}+2z\over L^{2}} & =  - \Lambda + {A^{2} \over 4} + {B^{2} \over 2}, \cr
 - {z + 2 \over L^{2}} & =  \Lambda + {A^{2} \over 4}.}}
\gaugeqn\ and \einstein\ then determine the necessary values of the cosmological constant and the fluxes:
\eqn\solution{\eqalign{\Lambda & = - {z^{2} + z + 4 \over 2 L^{2}}, \cr
A^2 & = {2z(z-1) \over L^2}, \cr
B^2 & = {4 (z-1) \over L^2}.}}

We notice that reality of the fluxes requires $z \geq 1$.  The dual field theories may exhibit ``critical slowing down," but never
``critical speeding up."

\newsec{Two-Point Correlation Functions}

We now explore the boundary observables defined by the theory in the bulk \metric\ by generalizing the usual holographic dictionary.
We specify henceforth to the case $z=2$, for illustration.
There are a variety of boundary
observables one could consider. We focus on the simplest possibility, two point correlation functions of scalar operators, in the present note.
After discussing generalities about the solutions for a bulk scalar field, we calculate the two-point function for a marginal operator (dual to a
massless scalar) in detail in \S 3.2.  We briefly discuss the general results for operators dual to massive scalars in \S 3.3.
We note that because the symmetries in our theories are much less constraining than the full conformal group, the general two-point function includes an unknown scaling
function of ${\bf x}^2/|t|$, and is not determined entirely by symmetry alone; so the two-point function already contains more non-trivial information than just a scaling
dimension.

\subsec{Real scalar field}

Consider a free, real scalar field $\phi$ in the Euclidean version of the background metric \metric.
We work in the coordinate $u = 1/r$ so that the boundary is located at $u=0$
\eqn\newmet{ds^2 = L^2 \left({1\over u^4 } ~d\tau^2 + {1\over u^2} ( d x^2 + d y^2 + d u^2 ) \right)~.}
The scalar has an action,
\eqn\scalaraction{S[\phi] = {1 \over 2} \int d^4 x~ \sqrt{g} \left( g^{\mu \nu} \partial_{\mu} \phi \partial_{\nu} \phi + m^{2} \phi^{2} \right).}
In a string construction such scalars could arise from the
moduli of the compactification manifold, from Kaluza-Klein modes of the metric and p-form fields, or from excited string states. The metric fluctuation $u^2 \delta g_{x y}$, among other things, also satisfies the same equation of motion as a bulk scalar with $m^2=0$ along the radial and temporal directions.

The field equation for $\phi$ is,
\eqn\phifieldeqn{\partial^{2}_u \phi - {3 \over u} \partial_u \phi + {u^{2}} \partial_{\tau}^{2}\phi +
(\partial_{x}^{2} + \partial_{y}^{2})\phi - {m^{2} L^{2} \over u^{2}} \phi = 0.}
Near the boundary a solution takes the asymptotic form,
\eqn\normnonnorm{\phi \sim c_{1} u^{\Delta_+} \phi_+(\tau,x,y) + c_{2} u^{\Delta_-} \phi_-(\tau, x,y),}
where $\Delta_{\pm}$, $\Delta_+ \geq \Delta_-$, are the two roots of the equation
\eqn\scalardimension{\Delta(\Delta - 4) =  m^{2} L^{2}.}

The requirement that the Euclidean action be finite imposes bounds on the allowed values of $\Delta$ and hence the mass of the field as in the case of the usual AdS/CFT correspondence \KlebanovTB.  For
\eqn\clearnorm{m^{2}L^{2} > -3}
there is a unique choice of boundary condition
\eqn\boundarycondition{\phi(u,\tau, x,y) \rightarrow u^{\Delta_+} (\phi(\tau, x,y) + O(u^2))} and via this choice, the scalar field is dual to an operator of dimension $\Delta_+ > 3$. In the window,
\eqn\window{-4 < m^2 L^2 < -3~,}
in addition to the above choice which remains valid, one can modify the Euclidean action to
\eqn\modifyaction{{1 \over 2} \int d^4x \sqrt{g}~ \phi (- \nabla^2 + m^2 ) \phi} by subtracting an infinite boundary term, so that the boundary condition
\eqn\modboundarycondition{\phi(u,\tau, x,y) \rightarrow u^{\Delta_-} (\phi(\tau, x,y) + O(u^2))} also leads to finite action. Thus there are two different quantizations for the scalar field in the range \window, and correspondingly there are two different non-Lorentz invariant fixed points on the boundary:
one with an operator of dimension $\Delta_+>2$, the other with an operator of dimension $1 < \Delta_- <  2$.
For $m^2 L^2 < -4$, the theory has a real instability; this is the analogue of the Breitenlohner-Freedman bound \refs{\BreitenlohnerBM} for the model in 2+1 dimensions
and with $z=2$
that we are studying.  The extension of these results to general spacetime dimension and dynamical exponent is straightforward.

\subsec{Correlation functions of marginal scalar operators}

As in standard AdS/CFT, boundary correlators are given by the value of the renormalized bulk action for specified boundary values of the bulk field (up to
cut-off dependent field renormalizations). To illustrate the
idea, consider first a free, real scalar field with $m=0$ for which
no field renormalization is necessary. This corresponds to an exactly marginal operator with scaling dimension four if we ignore gravitational
loop corrections. (Remember that the dual field theory lives
in $2+1$ dimensions with dynamical exponent $ z = 2$).

To compute the action, we find the boundary to bulk propagator $G(u,~ x;~ 0,~ x')$ that gives the nonsingular bulk field configuration $\phi(u,~x)$ for any smooth boundary value $\phi(0,~x)$ ($x$ denotes collectively $\tau, {\bf x} $)
\eqn\bbspace{\phi(u,~x)=\int d^3 x'~ \phi(0,~x')~ G(u,~x;~ 0,~x').}
The translational invariance in $\tau$ and ${\bf x}$ makes it easy to work in the Fourier space ($k$ denotes collectively $\omega, {\bf k}$) where \bbspace\ becomes
\eqn\bbfourier{\tilde \phi(u,~k) = \tilde G (u,~k)~\tilde \phi(0,~k)}
and $\tilde G (u,~\omega,~{\bf k})$ is the solution of \phifieldeqn with $m=0$ in Fourier space
\eqn\fouriereom{\partial^{2}_u \tilde G - {3 \over u} \partial_u \tilde G - (\omega^2~z^2 + |{\bf k}|^2) \tilde G = 0.}
with the boundary conditions
\eqn\bbb{\tilde G(0,~\omega,~{\bf k})=1}
and $\tilde G(u,~k)$ being finite as $u \rightarrow \infty$. These uniquely determine the propagator
\eqn\prop{\tilde G(u,~k) = e^{- | \omega | u^2/2}~\Gamma({|{\bf k}|^2 \over 4 | \omega |} + {3 \over 2})~ U({|{\bf k}|^2 \over 4 | \omega |}- {1 \over 2},~-1,~| \omega | u^2) .}
where $U(a,~b,~z)$ is the confluent hypergeometric function of the second kind. Note that $\tilde G(u, ~k)$ vanishes as $u \rightarrow \infty$.

By standard integration by parts, an on-shell bulk action is determined by the values of the field on the boundary
\eqn\onshell{\eqalign{S[\phi] & = \int d^{3}x \int_{\epsilon}^{\infty}d u \left(- \phi \partial_{\mu} \sqrt{g} g^{\mu \nu} \partial_{\nu} \phi + \partial_{\mu} ( \sqrt{g} g^{\mu \nu} \phi \partial_{\nu} \phi)\right) \cr
& = \int d^{3}x \left[ \sqrt{g} g^{u u} \phi \partial_u \phi~ \right]^{\infty}_{\epsilon} \cr & = \int d^{2} {\bf k} d\omega \left(\phi (0,~{\bf k},~\omega)~ {\cal F}({\bf k}, \omega) ~ \phi(0,~-{\bf k},~-\omega)  \right).}} We have cut off the whole space at $u = \epsilon$ to regulate the bulk action.  The `flux factor' ${\cal F}$ is
\eqn\fluxfactor{ {\cal F}({\bf k}, \omega) = \left[\tilde{G}(u,~{\bf - k},~-\omega) \sqrt{g} g^{uu}\partial_{u} \tilde{G}(u,~{\bf k},~\omega)\right]^{\infty}_{\epsilon}.}
Since the propagator $\tilde{G}$ vanishes at $u=\infty$, ${\cal F}$ only
receives a contribution from the cutoff at $u=\epsilon$.  The momentum-space two-point
function for the operator ${\cal O}_{\phi}$ dual to $\phi$ is calculated by differentiating \onshell\ twice with respect to $\tilde \phi(0,~{\bf k},~\omega)$,
\eqn\twoptmom{\langle {\cal O}_{\phi}({\bf k}, \omega) {\cal O}_{\phi}({\bf -k}, -\omega)  \rangle = {\cal F}({\bf k}, \omega).}
Near $u=0$, $\tilde{G}$ has the expansion,
\eqn\Gexpansion{\eqalign{\tilde{G} =  & 1 - {{\bf k}^{2} u^{2} \over 4} + {u^{4} \over 64}[3 {\bf k}^{4} - 20 \omega^{2} + 2 \gamma(4 \omega^{2} - {\bf k}^{4})  \cr
& + 8 {\bf k}^{2} | \omega | + (4 \omega^{2} - {\bf k}^{4})\log(u^{4} \omega^{2})+ 2 (4\omega^{2} - {\bf k}^{4})\psi({3 \over 2}+ {{\bf k}^{2} \over 4 | \omega |}) ] + {\cal O}(u^{6}),}}
where $\psi(x) = \Gamma'(x)/\Gamma(x)$ is the digamma function and $\gamma \approx .577$ is the Euler-Mascheroni constant.  Plugging the expansion \Gexpansion\ into \fluxfactor\ we pick out the leading non-polynomial piece in either ${\bf k}$ or $\omega$. This gives the correlation function as we take the cut-off $\epsilon$ to $0$
  \eqn\twoptmomresult{\eqalign{\langle {\cal O}_{\phi}({\bf k}, w) {\cal O}_{\phi}({\bf -k}, -w)  \rangle = & -{L^2 \over 2} {\bf k}^2 |\omega|- {L^2 \over 8} (4 \omega^{2} - {\bf k}^{4}) \log|\omega| \cr & - {L^2 \over 8} (4 \omega^{2} - {\bf k}^{4})  \psi({3 \over 2} + {{\bf k}^{2} \over 4 | \omega |}).}}
Specifically, the divergence arising as $\epsilon \rightarrow 0$ from the term proportional to $u^{2}$ in \Gexpansion\ is
removed via local boundary terms \BianchiKW, and the terms proportional
to $u^{4}$ in the first line of \Gexpansion\  give
rise to uninteresting contact terms in spacetime.
Terms ${\cal O}(u^{6})$ and higher vanish as the cutoff is removed $\epsilon \rightarrow 0$.

Terms of order $u^4$ in the second line of \Gexpansion\ are the only contributors to \twoptmomresult. Interestingly, among the three terms in \twoptmomresult, only the last one gives rise to correlations between points with both spatial and temporal separation.  The first two contribute terms localized in space.  The existence of such spatially localized terms is forbidden in Lorentz invariant theories, and may well be related to the ``ultra-local" behavior observed in \GVS.  Note also that \twoptmomresult\ has the correct scaling behavior for the two-point correlator of a dimension four operator.  Its momentum and frequency dependence is complicated, however.

One can understand the large distance fixed time behavior of the correlation function by Fourier transforming \twoptmomresult\ to position-space.  Only the piece containing the digamma function contributes in this regime, and its Fourier transform can be computed by utilizing the expansion
\eqn\psiexpansion{\psi(x) = - \gamma - {1\over x} - \sum_{n=1}^{\infty} \left({1 \over x+n} - {1 \over n}\right).}
It suffices to quote that the large distance behavior of the two point function is just a simple power law
\eqn\twoptspace{ \langle {\cal O}_{\phi}({\bf x},~t) {\cal O}_{\phi}({\bf 0}, 0)  \rangle \rightarrow {const \over |{\bf x}|^8},\,\,\,\,\,|{\bf x}| \rightarrow \infty .} This
is interesting because a priori, one might have expected further suppression of the power law result by some scaling function of ${\bf x}^2/|t|$ (e.g. $e^{-{{|\bf x|^2} \over |t|}}$, which would result in an ultra-local equal time two-point function)
could have arisen,
since it is consistent with the symmetries of the theory.
We note that the correlators of the simplest scaling operators in the free Lifshitz theory \lif\ also exhibit pure power law decay at large spatial separation \Fradkin.

\subsec{Correlation functions of generic scalar operators}

It is a direct generalization of the above to compute the two point functions for operators which are dual to massive scalar fields, complicated only by additional renormalization effects. Specifically, to define the boundary to bulk propagator, we need to cut off the space and put the boundary at $u= \epsilon$. The propagator is now a solution to
\eqn\fouriereom{\partial^{2}_u \tilde G - {3 \over u} \partial_u \tilde G - (\omega^2~z^2 + |{\bf k}|^2 + {m^2 L^2\over z^2}) \tilde G = 0 } satisfying the boundary conditions
\eqn\bbbm{\tilde G (\epsilon, {\bf k}, \omega) = 1} and $\tilde G$ nonsingular for $u> \epsilon$.

For generic values of $m$, this determines
\eqn\bbpm{\tilde G(u,~k) \propto e^{- {1 \over 2} |\omega| u^2} u^{
 2 + \sqrt{4 + m^2 L^2}}~ U({| {\bf k}|^2 \over 4 | \omega |} + {1 \over 2}  + \sqrt{1 + {m^2 L^2 \over 4}} , 1 + \sqrt{4 + m^2 L^2}, | \omega | u^2 )} with the proportionality constant set by \bbbm. Now the leading non-polynomial term in $\tilde G$ that contributes to the flux factor would come at ``order'' $u^{2 +  \sqrt{4 + m^2 L^2}}$
 \eqn\Gmexpansion{\eqalign{\tilde G (u, ~k) & = \left({u \over \epsilon} \right)^{2- \sqrt{4 + m^2 L^2}} [1+...+ (|\omega| u^2)^{ \sqrt{4 + m^2 L^2}}  \cr &  \times {\Gamma(-\sqrt{4+ m^2 L^2}) \over \Gamma(\sqrt{4+ m^2 L^2})} {\Gamma({|{\bf k}|^2 \over 4 |\omega|} + {1 \over 2} + \sqrt{1 + {m^2 L^2 \over 4}}) \over \Gamma ({|{\bf k}|^2 \over 4 |\omega|} + {1 \over 2} - \sqrt{1 + {m^2 L^2\over 4}})}+... ].}} The two point function in momentum space is again given by the flux factor, and after throwing away divergent and vanishing pieces as we take $\epsilon \rightarrow 0$, we find
\eqn\twoptm{\eqalign{\langle {\cal O}({\bf k}, \omega) {\cal O}({\bf -k}, -\omega)  \rangle & = - (2+ \sqrt{4 + m^2 L^2})~L^2~ |\omega|^{\sqrt{4 + m^2 L^2}} \cr & \times {\Gamma(-\sqrt{4+ m^2 L^2}) \over \Gamma(\sqrt{4+ m^2 L^2})} {\Gamma({|{\bf k}|^2 \over 4 |\omega|} + {1 \over 2} + \sqrt{1 + {m^2 L^2 \over 4}}) \over \Gamma ({|{\bf k}|^2 \over 4 |\omega|} + {1 \over 2} - \sqrt{1 + {m^2 L^2\over 4}})}.}} Again \twoptm\ has the correct scaling behavior for the correlator of an operator of dimension $2 + \sqrt{4 + m^2 L^2}$, but it also involves
additional momentum and frequency dependence\foot{Note that for non-generic values of the scalar mass
$m^2 L^2 = n^2 - 4$ with $n$ being non-negative integers, the above formula \twoptm\ does not apply.}.

\newsec{Holographic RG flow between $z=2$ Lifshitz fixed points and $z=1$ conformal field theories}

It is natural to expect that theories with anistropic scale invariance can flow, under relevant perturbation, to fixed points which enjoy full conformal invariance.
For instance, for the toy model \lif, perturbation by $- (\nabla \phi)^2$ induces RG flow to the free massless scalar field theory, which is a conformal field theory.

Here, we find that our gravitational duals of Lifshitz-like theories come with relevant perturbations that induce a similar flow.
We will look for solutions which represent holographic RG flow from the $z=2$ Lifshitz-like fixed point in the UV (at large $r$) towards an $AdS_4$-like spacetime in the IR (small $r$).
A metric ansatz which is sufficiently general to capture this flow is:

\eqn\metricflow{ ds^2 = L^2 \left(- r^4 f(r)^2 dt^2 + r^2 (d x^2 + d y^2) + g(r)^2 d r^2/r^2 \right).}

If $f(r)=g(r) \equiv 1$, this is the spacetime dual to the $z=2$ fixed point. If $f(r)$ is inversely proportional to $r$, and $g(r)$ is constant, it is $AdS_4$.
Introduce again the non-coordinate basis of 1-forms in which the metric simplifies to constant Minkowskian form $\eta_{\mu \nu}$:

\eqn\thetaflow{\eqalign{ \theta_t &=L r^2 f(r) d t \cr \theta_x &=L r d x \cr \theta_y &=L r d y \cr \theta_r &= L g(r) d r/r}}

In terms of this basis, the two and three form fluxes are

\eqn\twoformflow{F_{(2)} = {2 \over L} h(r) \theta_r \wedge \theta_t}
\eqn\threeformflow{F_{(3)} = {2 \over L} j(r) \theta_r \wedge \theta_x \wedge \theta_y.} The advantage of using the $\theta$ basis is that the fluxes remain nonsingular as long as
the coefficient functions $h(r)$ and $j(r)$ are finite. In addition to the fluxes, we also have a negative cosmological constant $\Lambda = -5/L^2$. The $z=2$
Lifshitz fixed point has $h(r)=j(r)\equiv 1$ and the IR $AdS_4$ has $h(r)=j(r)\equiv 0$

The equations of motion for the fluxes and the metric, after proper massaging, give:

\eqn\fghjflow{\eqalign{2 r f'/f  & = (5-h^2+j^2) g^2-5 \cr
r g' & = {1 \over 2} g^3 (h^2+j^2-5) + {3 \over 2} g \cr
r h' & = 2 g j -  2 h \cr
r j' & = 2 g h + {1\over 2} j + {1\over 2} j g^2 (h^2-j^2-5)~.}} The counting of
the equations works as follows: the two and three form fluxes each satisfy a single equation of motion, and the Einstein equation for our background \metricflow, \twoformflow, and \threeformflow\
results in only two independent equations. It takes some algebra to
show the latter fact. It is easy to check that the $z=2$ Lifshitz fixed point and $AdS_4$ are the only two fixed points of the above flow equations.

Note the relation between the above four equations. The last three form a closed set of ODE's for the three functions $g(r), h(r)$ and $j(r)$. Given any initial values for them at some fixed $r_0$, we can determine their behavior as functions of $r$. Then by the first equation, we can determine the behavior of $f(r)$. Note that the initial value of $f$ is not a physical parameter, but a gauge choice. It can be changed by a global rescaling of the $t$-coordinate.

Linearization near the fixed points gives information about the attractiveness of the fixed points. Around the $AdS_4$ fixed point, the flow matrix has eigenvalues $(-3, -3.13, 0.13)$. The first corresponds to a flow along the direction $\p_g$ away from the $AdS_4$ fixed point as we flow into the IR, the second and third describe flow away and towards the fixed point along two orthogonal directions inside the plane of $\{ \p_h, \p_j \}_{AdS_4}$. Around the Lifshitz fixed point, the flow matrix takes the Jordan normal form

\eqn\flowmatrixlif{\pmatrix{0 & 0 & 0 \cr 0 & -4 & 1 \cr 0 & 0 & -4}} in for example the basis \eqn\basislif{\pmatrix{{3 \over 8} \p_h + {1 \over 4} \p_j \cr \cr {1 \over 8} \p_h-{1 \over 4} \p_j \cr \cr \p_g-{1\over 2} \p_h}.} There are two relevant directions (with logarithmic mixing between them) and one marginal direction around this fixed point.

One can numerically solve the system \fghjflow\ using the ``shooting" technique.  Since the AdS fixed point has relevant perturbations, it is hard to hit on the nose by varying the UV perturbation.
Instead, we start with the AdS fixed point close to $r=0$ and try to hit the $z=2$ critical point by integrating the flow out towards $r = \infty$; this is much easier to do, because the UV fixed point has only relevant and marginal perturbations, and is hence (almost) UV attractive.  One can find good flows that asymptote between the two fixed points in this way; an example is displayed in Figure 1 below.

\bigskip
\centerline{\epsfxsize 2.3truein\epsfbox{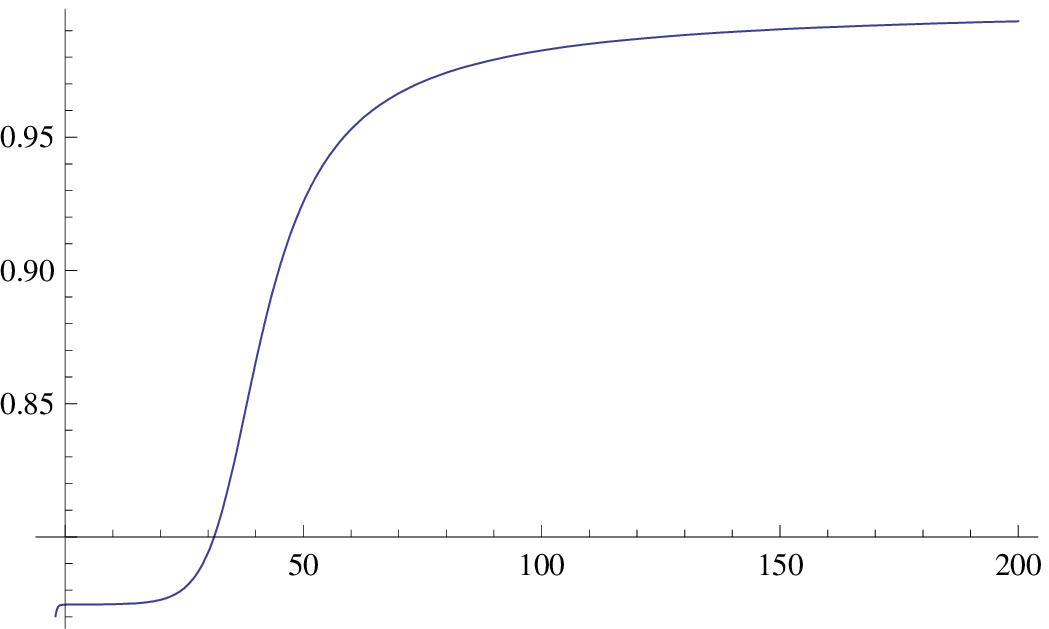} {$\,\,\,$} \epsfxsize 2.3truein\epsfbox{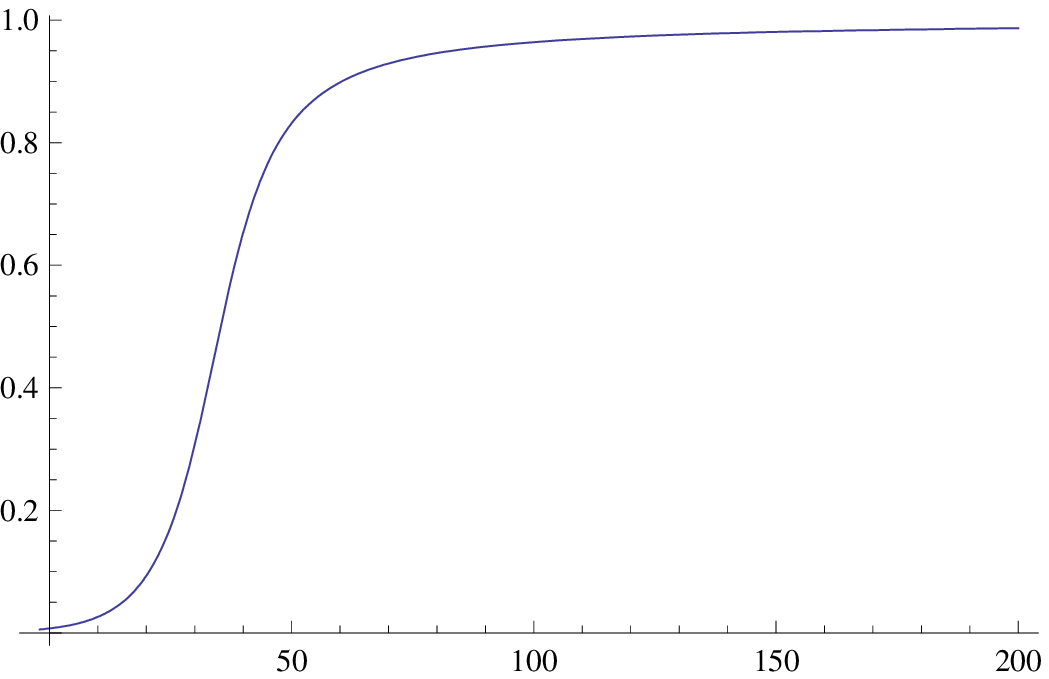} {$\,\,\,$} \epsfxsize 2.3truein\epsfbox{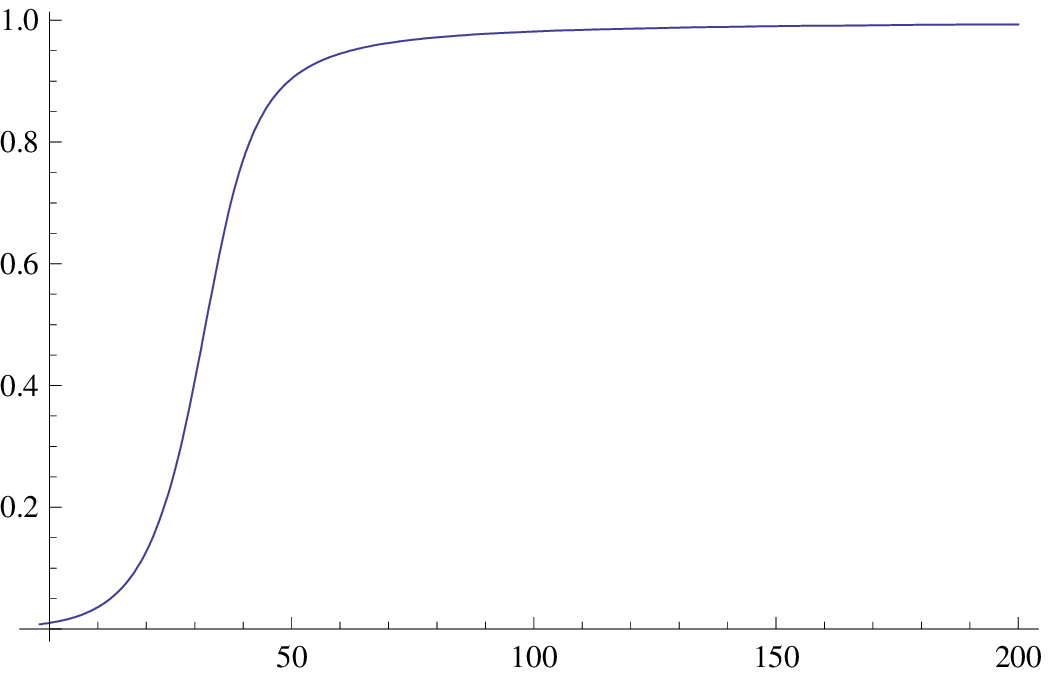}}
\noindent{\ninepoint
\baselineskip=2pt {\bf Fig. 1.} {Plotted are respectively $g(r)$, $h(r)$ and $j(r)$ as functions of $log(r)$. Their initial values are
set at $r=1$ to be respectively $\sqrt{3 \over 5} $, $0.00728$ and $0.01$, 
i.e. small deviations away from the $AdS_4$ fixed point along the unique irrelevant direction. As the figure
shows, all three functions asymptote to $1$ at large $r$. Numerical integration of the system \fghjflow\ suggests that
the marginal direction at the Lifshitz fixed point becomes relevant at the nonlinear level.}}

\newsec{Future directions}

There are many directions for further research.  They range from straightforward extensions of this work to somewhat
speculative but intriguing possibilities:

\noindent
$\bullet$  It would be interesting to compute more complicated observables
in these spacetimes, including higher-point correlation functions and
Wilson loops.

\noindent
$\bullet$ By adding bulk ``probe" fields and varying the parameters in
their Lagrangian, it should be possible to find interesting quantum phase
transitions in this system.

\noindent
$\bullet$ The embedding into a full string theory construction of our
solutions is left as work for the future.  We anticipate that this should
be possible using standard techniques of flux compactification.

\noindent
$\bullet$ A better understanding of the global and causal structure of the spacetime \metric\ would also be desirable.  In particular, it is necessary to understand the physics behind the peculiar behavior
at $r=0$ in more detail.  This kind of behavior has been observed before in string theoretic
brane systems \Gary.
It is possible that in this case, the behavior is related to the strong infrared singularities which one might expect
in interacting Lifshitz-like theories.\foot{We thank S. Dubovsky for discussions of the peculiar
infrared behavior of these theories.}

\noindent
$\bullet$ Our solutions have correlation functions that do not exhibit
ultra-locality at zero temperature.  However, in the work of \GVS,
it was found that the Lifshitz fixed point has ultra-local correlators
at finite temperature $T$, 
but not at vanishing $T$.  (Ultra-locality was also observed
in the $SU(2)$ gauge theories discussed by \Chetan).
It is therefore interesting
to construct finite $T$ analogues of our spacetime (which presumably
amounts to finding black hole solutions), and see whether the correlators
exhibit some of the same behavior discussed in \GVS.

\noindent
$\bullet$ Finally, in any application of gauge/gravity duality to a strongly
coupled system, it is important to understand where the large $N$
degrees of freedom in the field theory are supposed to arise.  In many
field theories of interest to condensed matter theorists, there is of course
no large $N$ and no 't Hooft-like expansion.  However, the results of
\Chetan\ suggest that novel 3D gauge theories with $SU(N)$ gauge group may
well have Lifshitz-like fixed points (though they focused on the case
$N=2$).  It seems worthwhile to try and  generalize their results to the planar
limit of large $N$ theories; this would potentially
give a direct point of contact
between a class of novel field theories, and gravity in spacetimes of the
sort we studied here.

\eject


\centerline{\bf{Acknowledgements}}
\medskip
We are happy to thank A. Adams, J. Cardy, E. Fradkin, P. Horava, J. Maldacena,
J. McGreevy, G. Moore, C. Nayak, S. Sachdev, S. Shenker and
E. Silverstein for stimulating discussions.  We especially appreciate J. Maldacena's absolutely valuable comment
about the calculation of correlators in this spacetime.
We are also grateful to M. Amin for help with numerical methods.  M.M. thanks 
C.-Y. Hou, C. Laumann, S.
Parameswaran, and A. Rahmani for computer assistance after catastrophic HD failure.
S.K. would like to acknowledge the kind hospitality
of the Kavli Institute for Theoretical Physics, the Aspen Center for Physics, and the Institute for Advanced Study
at various points during this work. X.L. would like to acknowledge the kind hospitality of the SITP during the major phase of this collaboration.
M.M. thanks the Les Houches School of Physics for hospitality during the completion of this work.
This research was supported in part by the Stanford Institute for Theoretical Physics, the NSF under grant PHY-0244728, and the DOE under contract DE-AC03-76SF00515.  M.M. was also supported by an ARCS Fellowship. Research at the Perimeter Institute for Theoretical Physics is supported in part by the government of Canada and through NSERC and by the province of Ontario through MRI.

\listrefs
\end